\documentclass[twocolumn,superscriptaddress,aps,prl]{revtex4-2}

\usepackage{graphicx}
\usepackage{dcolumn}
\usepackage{bm}
\usepackage{subfigure}
\usepackage{amsmath}
\usepackage{mathrsfs}
\usepackage{amssymb}
\usepackage{setspace}
\usepackage{array}
\usepackage{multirow}
\usepackage{float}
\usepackage{flushend}
\usepackage{footmisc}
\usepackage{subfigure}

\usepackage[pdfstartview=FitH,
CJKbookmarks=true,
colorlinks,
linkcolor=blue,
anchorcolor=blue,
citecolor=blue,
urlcolor=blue,
]{hyperref}

\usepackage{footmisc}

\renewcommand{\footnoterule}{
	\kern -4pt  
	\hrule width 0.18\linewidth height 0.6pt
	\kern 12pt 
}

\usepackage{footmisc}

\usepackage{braket}

\begin{document}

\preprint{APS/123-QED}

\title{Robustness of quantum many-body scars in the presence of Markovian bath}

\author{Xiang-Ping Jiang}
\affiliation{School of Physics, Hangzhou Normal University, Hangzhou, Zhejiang 311121, China}

\author{Mingdi Xu}
\affiliation{School of Physics, Nankai University, Tianjin 300071, China}

\author{Xuanpu Yang}
\affiliation{School of Physics, Nankai University, Tianjin 300071, China}

\author{Hongsheng Hou}
\affiliation{School of Physics, Hangzhou Normal University, Hangzhou, Zhejiang 311121, China}

\author{Yucheng Wang}
 
\affiliation{Shenzhen Institute for Quantum Science and Engineering,
Southern University of Science and Technology, Shenzhen 518055, China}
\affiliation{International Quantum Academy, Shenzhen 518048, China}
\affiliation{Guangdong Provincial Key Laboratory of Quantum Science and Engineering, Southern University of Science and Technology, Shenzhen 518055, China}

\author{Lei Pan}%
\email{panlei@nankai.edu.cn}
 
\affiliation{School of Physics, Nankai University, Tianjin 300071, China}


\begin{abstract} 

A generic closed quantum many-body system will inevitably tend to thermalization, whose local information encoded in the initial state eventually scrambles into the full space, known as quantum ergodicity. A paradigmatic exception in closed quantum systems for strong ergodicity breaking is known as many-body localization, where strong disorder-induced localization prevents the occurrence of thermalization. It is generally recognized that a localized quantum system would be delocalized under dissipation induced by the environment. However, this consequence recently has received challenges where an exotic dissipation-induced localization mechanism is proposed, and transitions between localized and extended phases are found. In this Letter, we promote this mechanism to systems for weak ergodicity breaking hosting quantum many-body scars (QMBS). We find that the system relaxes to a steady state dominated by QMBS, and the dissipative dynamics exhibit dynamic revivals by suitably preparing an initial state. We point out an experimental realization of the controlled dissipation with a cold atomic setup. This makes the signature of ergodicity breaking visible over dissipative dynamics and offers potential possibilities for experimentally preparing stable QMBS with associated coherent dynamics.

\end{abstract}

\maketitle

\textit{\color{blue}Introduction.} Within the framework of statistical mechanics, generic thermodynamic closed systems relax to thermal equilibrium independent of initial conditions \cite{Gogolin,Rigol2016}. The thermalization of closed quantum systems can be interpreted via the eigenstate thermalization hypothesis (ETH) \cite{ETH1,ETH2,ETH3} which states that the eigenstate itself is thermalized.  
However, there exist counter examples violating above paradigm such as disordered systems. A prototype is many-body localization (MBL) \cite{MBL1,MBL2} which is a consequence of the balance between many-body interactions and disorder-induced localization. The MBL emerges as a paradigm for strong ergodicity breaking \cite{StrongETH1,StrongETH2,StrongETH3,StrongETH4,StrongETH5} and the corresponding quench dynamics can persist to long lasting 
times, providing an alternative to distinguish those systems with thermalizing dynamics. However, realistic quantum systems are seldom completely isolated from external environment, raising the question of whether MBL can persist in the presence of dissipation over extended timescales.

Intuitively, any localized system will be inevitably thermalized when it couples a thermal environment that the local information of initial state scrambles in the whole space. Previous theoretical studies and experimental measurements about dissipative dynamics of the MBL in presence of environment showed that dissipation spoils the MBL and induces ergodicity eventually with a slow relaxation dynamics \cite{OpenMBL_1,OpenMBL_2,OpenMBL_3}. This is physically evident that any local dephasing will drive the system into maximally mixed state (infinite temperature state) independent on choice of the system Hamiltonian despite a MBL exhibits the stretched exponential relaxation signature differ from an ergodic system. 
However, above analysis is valid for cases that dissipative operators are hermitian or local (on-site) operators.  Recently, several works re-investigated the fate of localization properties in disordered systems under a special type of \textit{quasi-local} dissipation operators , and found that the system can be driven into new steady states exhibiting signatures of localization \cite{Yusipov,Yusipov2,WYC_PRL,WYC_MBL,PL_mosaic,PL_3D} and a dissipation induced localized-extended transition was proposed \cite{WYC_PRL}.
This provides a feasible route towards preparing non-thermal quantum states violating the ETH via engineered dissipation. 

In the last few years, a new class of paradigms that violate the ETH has emerged \cite{scar1,scar2,Fragment,Buca2023}. Among these, quantum many-body scars (QMBS) as an example of weak ergodicity breaking have gained significant attention. QMBS were first discovered in Rydberg atom arrays, where persistent revivals are observed for certain special initial states, while the majority of initial states exhibit rapid relaxation \cite{Rydberg_Exp1,Rydberg_Exp2}.

The long-term coherent dynamics observed in these systems stem from the quantum many-body scarring mechanism \cite{scar1,scar2}, which is defined by a small subset of eigenstates, known as quantum many-body scars (QMBS), that violate the eigenstate thermalization hypothesis (ETH). In contrast to strong ergodicity-breaking phenomena, such as many-body localization (MBL), where the entire spectrum deviates from the ETH, QMBS represent a measure-zero subset of the spectrum. In the thermodynamic limit, the number of QMBS scales exponentially smaller than the total dimension of the Hilbert space, rendering them a rare yet fascinating feature of non-integrable quantum systems.

This could make one expect that QMBS do not show an experimentally observable dynamic feature since they are buried in the energy spectrum.
However, those systems containing QMBS as equally 
spaced towers in the spectrum can result in dynamical revival from special initial states, that was observed in Rydberg atom experiments. Although there have been extensive research on QMBS in closed systems \cite{scar_Review1,scar_Review2,scar_Review3}, the study of QMBS in the framework of open quantum systems and non-hermitian physics is still in its infancy \cite{NH_scar1,NH_scar2,Open_scar1,Open_scar2}.

In this Letter, we explore the impact of dissipation on systems with QMBS and explore whether dissipation can drive a system to steady states exhibiting structures of QMBS instead of destroying them and then ascertain if the persistent revivals can be witnessed in corresponding dissipative dynamics for particular initial states. We demonstrate that by utilizing a set of quasi-local dissipation operators acting on a pair of neighboring sites, systems can be driven into a specific
steady state consisting mainly of QMBS. We confirm the above finding through three representative models containing QMBS, i.e., the PXP model with approximate QMBS, the toy model with exact QMBS embedding within nonintegrable models \cite{Toymodel}, and the generalized $J_1-J_2$ model where QMBS can be analytically solved by spectrum generating algebra \cite{J1_J2,SGA,Liang,Tang,Popov}. Moreover, we provide an experimental proposal to realize such quasi-local dissipation operators and observe related dissipative scarring dynamics based on current techniques about Rydberg atomic simulators.\\

\begin{figure}[!ht]
	\includegraphics[width=8.0cm]{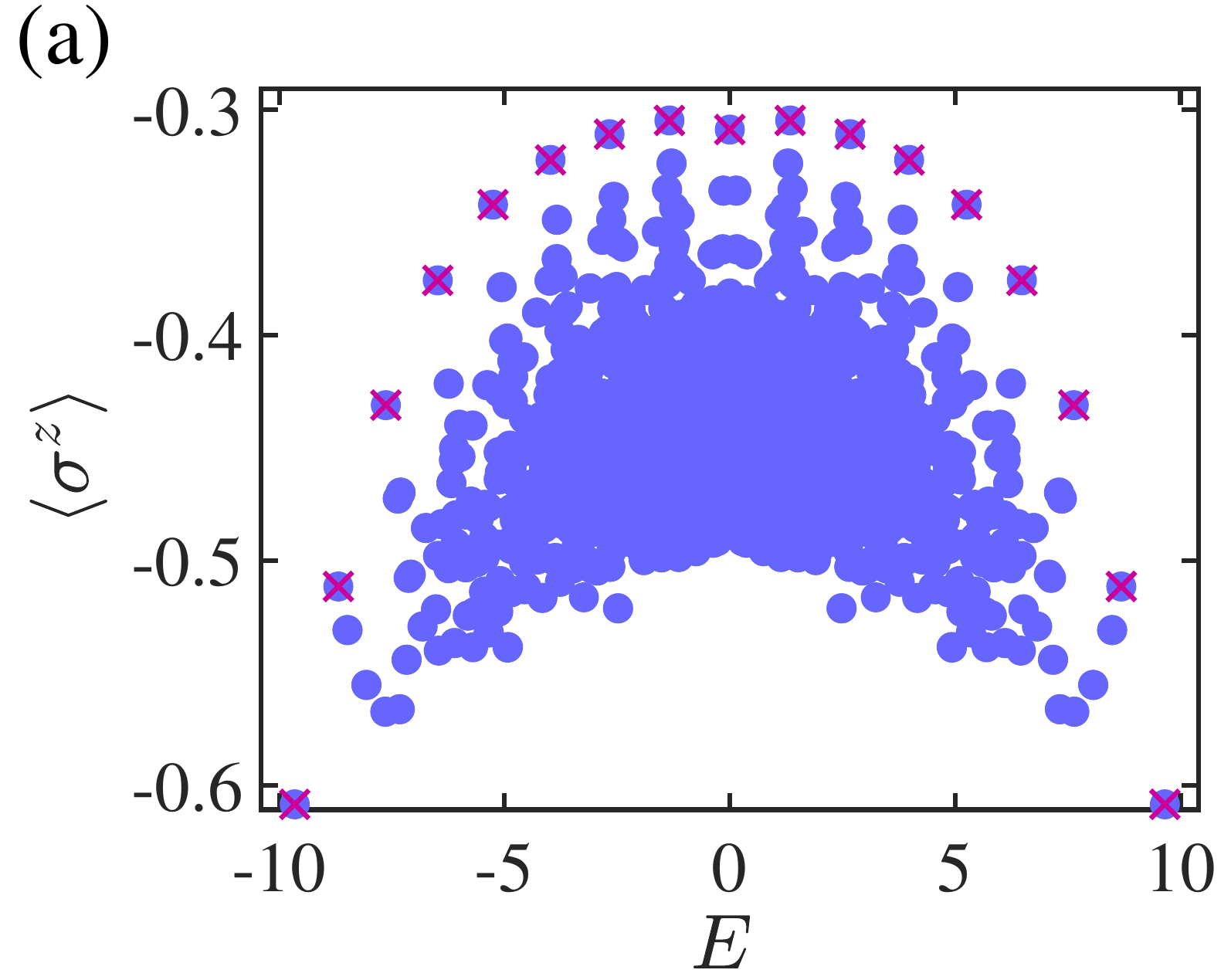}
	\includegraphics[width=8.0cm]{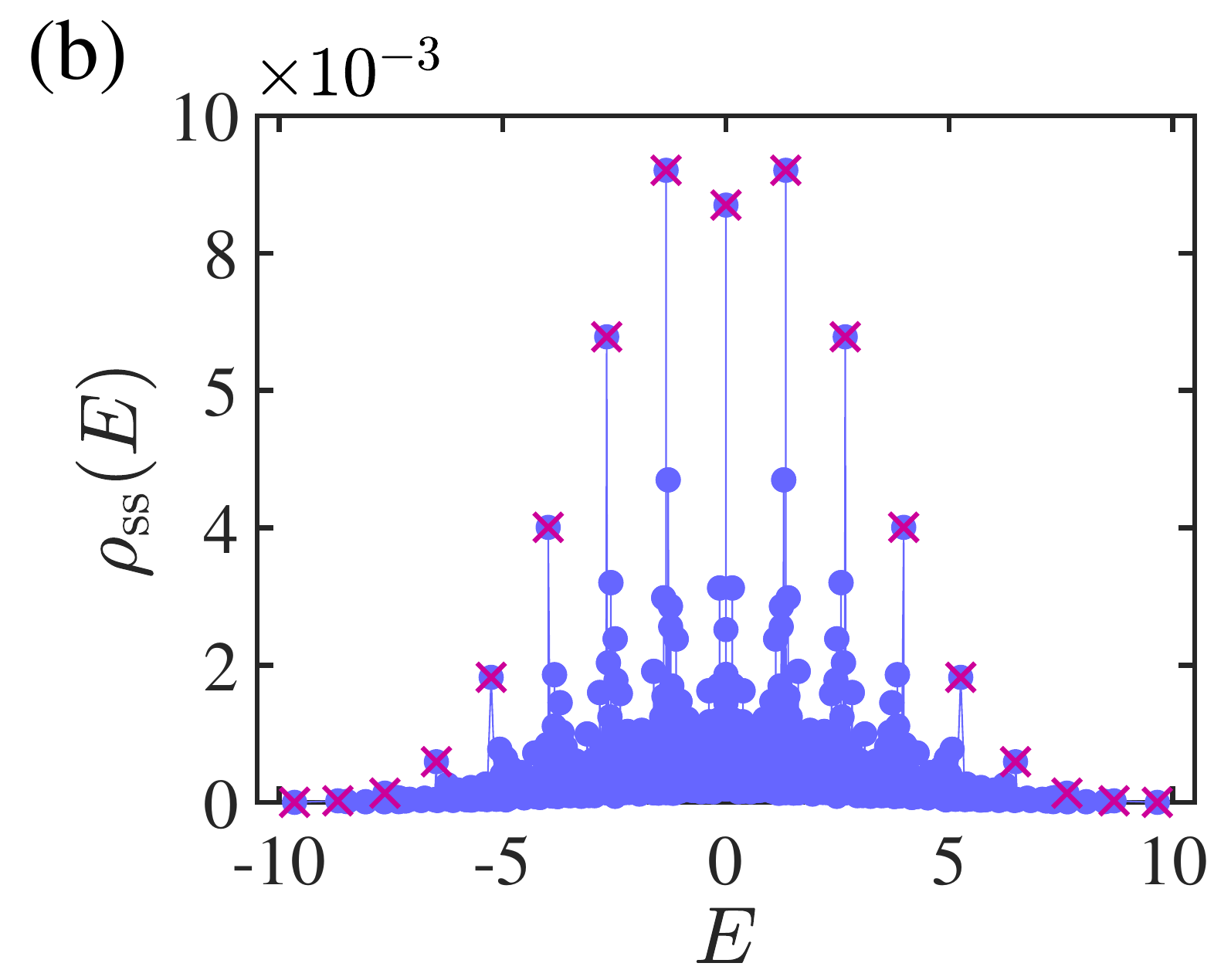}
	\caption{(a) The expectation value of mean magnetization $\langle \sigma^z\rangle=\sum_{j}\langle \sigma_j^{z}\rangle/L$ in the PXP model as a function of eigenenergy $E$. Most of the points are concentrated in the thermal bulk, but there are outermost eigenstates (indicated by crosses) as QMBS violating the ETH. 	(b) Distribution of the steady-state density matrix on eigenstates of the PXP model as a function of energy corresponding with (a). It is visible that QMBS have larger weights than thermal states in the vicinity of energy. Here, the system size is chosen as $L=16$, and the dissipative phase and strength are $\theta=\pi$ and $\gamma=0.1$, respectively.}
	\label{Fig1}
\end{figure}

\textit{\color{blue}Dissipative PXP model.} A prominent system containing QMBS is the PXP model, which serves as an effective model \cite{scar2} for the Rydberg atoms array. This model plays a pivotal role, and the corresponding Hamiltonian in one dimension is 
\begin{align}
H_{\mathrm{PXP}}= \sum_j  P_{j}\sigma_{j+1}^xP_{j+2}, \label{PXP_Ham}
\end{align}
where $P_{j}=\ket{0_j}\bra{0_j}=\frac{1-\sigma_{j}^z}{2}$ is the projector onto $\ket{0_j}$ state denoting no Rydberg excitation at the site $j$ where the basis on each site consists of $\ket{n_j}$ with Rydberg excitation number $n_j=0,1$ in which case the operator $\sigma_{j}^x=\ket{0_j}\bra{1_j}+\ket{1_j}\bra{0_j}$ ($\sigma_{j}^z=\ket{1_j}\bra{1_j}-\ket{0_j}\bra{0_j}$) represents the Pauli matrix in $x$ ($z$) direction.
The PXP Hamiltonian \eqref{PXP_Ham} has the constraint stemming from the Rydberg blockade effect \cite{blockade1,blockade2,blockade3,blockade4} that each spin can freely rotate only if both its neighbors are no Rydberg excitation $\ket{0_j}$, while it is frozen otherwise. Despite its simplicity, the PXP Hamiltonian is nonintegrable, with its level statistics approaching the Wigner-Dyson distribution in the thermodynamic limit. However, within its spectrum, a subset of nearly equidistant special eigenstates, known as quantum QMBS, forms a band with a tower-like structure spanning the entire spectrum. As illustrated in Fig.~\ref{Fig1}(a), these QMBS are visibly separated from the bulk of the spectrum. In the PXP model, QMBS are responsible for the periodic revivals observed in quench dynamics, with the revival period being approximately inversely proportional to the energy spacing between the QMBS.

The QMBS are generally fragile to perturbation \cite{Scar_perturbation1} or disorder \cite{Scar_perturbation2}. It is naturally expected that QMBS are also unstable against the interaction between the system and external environments. Suppose the Rydberg atom chain described by the PXP model is coupled to a Markovian bath, whose dynamics are determined by the following Lindblad master equation \cite{Lindblad1,Lindblad2}

\begin{align}
	\frac{d \rho(t)}{dt} = \mathcal{L}[\rho(t)] = -i [H_S, \rho(t)] + \mathcal{D}[\rho(t)]\label{Lindblad}
\end{align}
where $\rho(t)$ is the density matrix of the system, $H_S$ denotes the system Hamiltonian and $\mathcal{L}$ is referred to as the Liouvillian superoperator. The first part on right-hand side of the Lindblad equation \eqref{Lindblad} is the unitary evolution, and the second part represents dissipative evolution, which is written by
\begin{align}
	\mathcal{D}[\rho(t)] = \sum_{j} \gamma_{j}\left(2 O_{j} \rho O_{j}^{\dagger}-\Big\{O_{j}^{\dagger} O_{j}, \rho \Big\} \right).\label{Dissipator}
\end{align}
Here $\{,\}$ denotes the anticommutator, and $O_{j}$ is the dissipation operator or jump operator at the site $j$ with dissipation strength $\gamma_{j}$. As with the system Hamiltonian determining time evolution of a closed system, the dissipative dynamics of an open quantum system are governed by the Liouvillian superoperator. The corresponding solution is formally expressed as $\rho(t)=e^{\mathscr{L}t}\rho_0$ and the system reaches its steady state $\rho_{ss}= \lim_{t \to \infty}\rho(t)= \lim_{t \to \infty}e^{\mathscr{L}t}\rho_0$ for arbitrary initial state $\rho_0=\rho(0)$. The steady state is closely relevant to the choice of dissipation operators $O_{j}$, and a particular concern is to design the dissipation operators such that the steady-state serves as the required quantum state. This paper's central thesis is to make QMBS a steady state by using specific dissipation operators as simply as possible.

\begin{figure}[!ht]
	\includegraphics[width=8.0cm]{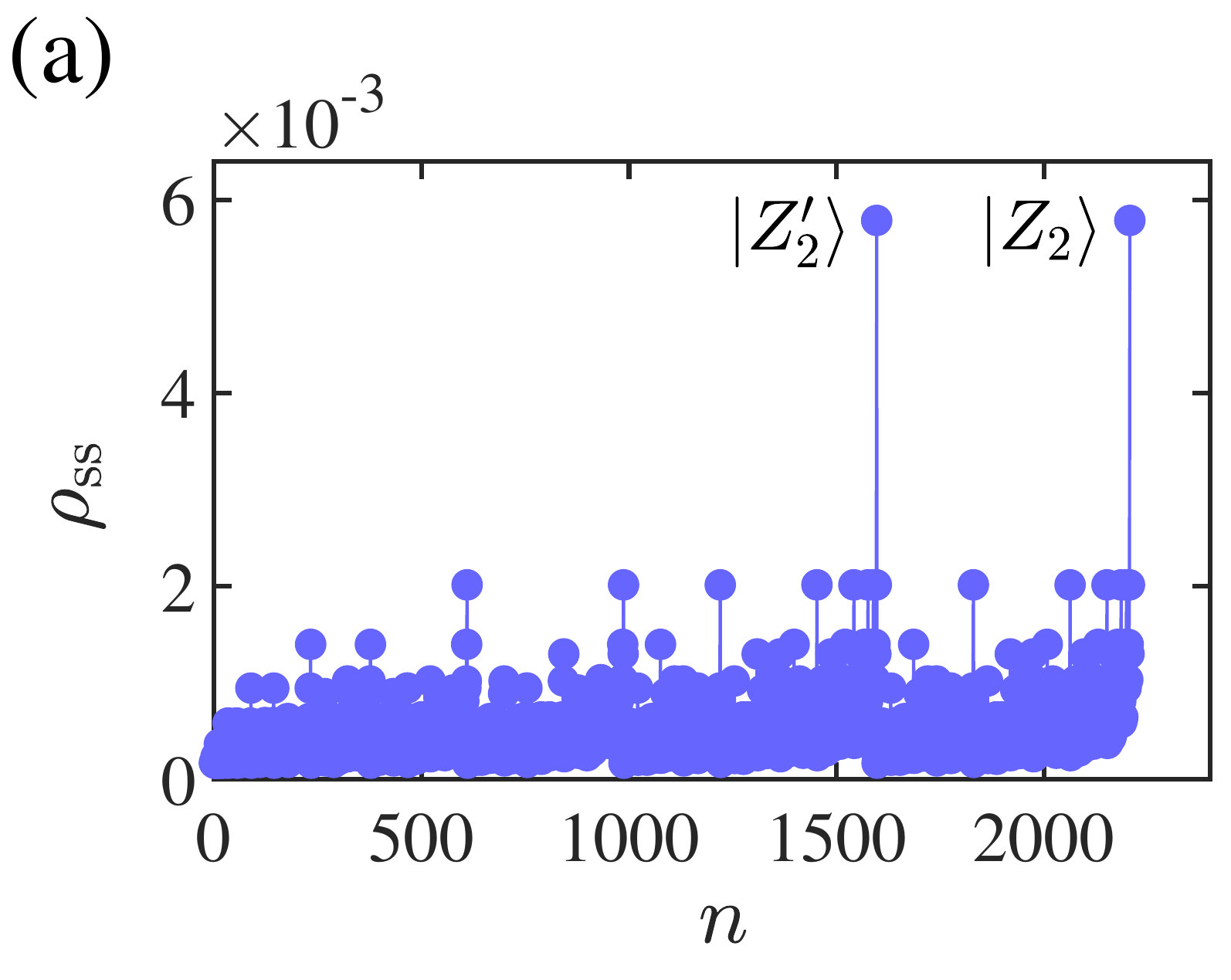}
	\includegraphics[width=8.0cm]{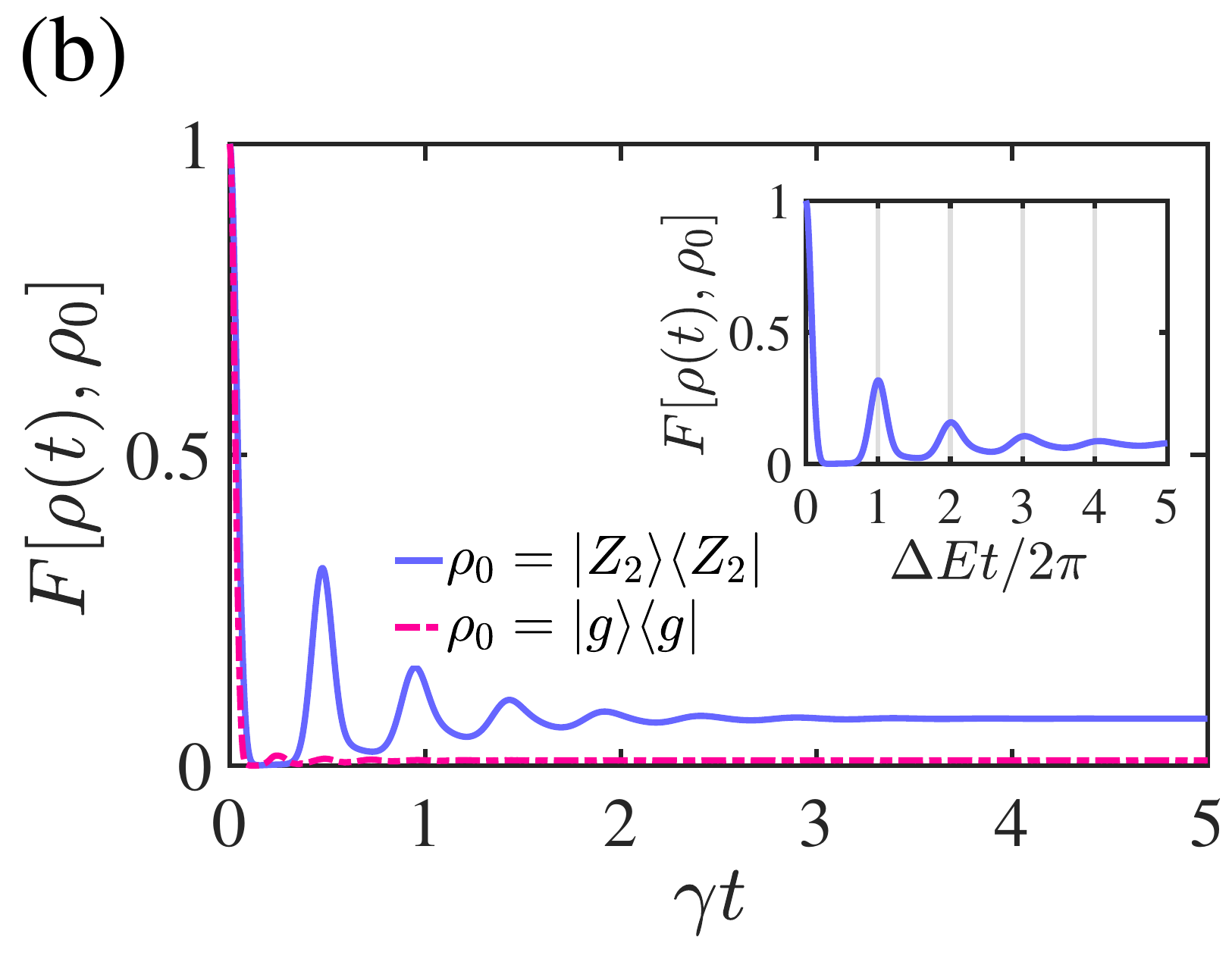}
	\caption{ (a) (a) Distribution of the steady-state density matrix in the local basis of the PXP model, highlighting two product states ($\ket{Z_2}$ and $\ket{Z_2'}$ ) with significantly large values. (b) Time evolution of the quantum fidelity for different initial product states. The system rapidly relaxes to negligible steady-state value (red dashed line) from the initial state $\ket{g}=\ket{000\cdots000}$ without Rydberg excitations. In contrast, for the $\ket{Z_2}$ initial state, the fidelity shows periodic oscillations before reaching a non-zero steady-state value, and the period is $T=2\pi/\Delta E$ where $\Delta E$ is the mean energy spacing of QMBS near the middle of the spectrum. The system size and dissipation parameters are identical to those used in Fig.\ref{Fig1}}
	\label{Fig2}
\end{figure}

Here we consider the following types of jump operators
\begin{align}
	\begin{split}
		O_{j} = \left( \sigma_{j}^++e^{i\theta}\sigma_{j + 1}^+ \right) 
		\left(  \sigma_{j}^- -e^{i\theta}  \sigma_{j+1}^-\right),
	\end{split}
	\label{O_j}
\end{align}
These operators were first introduced in Refs. \cite{Jump1,Jump2} to realize topological phases as steady with engineered dissipation. The physical implementation can be based on a Bose-Hubbard model in superconducting resonators array \cite{BHchain}, and also be proposed on optical Raman lattices \cite{WYC_PRL}. Physically, the above jump operator acts on a pair of nearest-neighbor sites and changes the relative phase between them but does not alter particle number. Moreover, it not only includes the local on-site dephasing term $n_j=\sigma_{j}^+\sigma_{j}^-$ but also the (quasi-local) hopping term  $e^{i\theta}\sigma_{j}^+\sigma_{j+1}^--e^{i\theta}\sigma_{j+1}^+\sigma_{j}^-$ which is crucial to our objectives. Now we consider such dissipation operators \eqref{O_j} act on the PXP model, and the corresponding steady state can be obtained from the zero-mode of Liouvillian superoperator. In Fig.\ref{Fig1}(b), we plot the steady-state distribution on eigenstates of the PXP Hamiltonian, where a structure is clearly visible that the QMBS obtain larger weights than those states in the vicinity of energy. A key physical significance of QMBS in the PXP model lies in their high overlap with specific product states ($\ket{Z_2}$ and $\ket{Z_2'}$), which are experimentally accessible in Rydberg atom setups. To uncover the intrinsic structure of QMBS, we calculate the distribution of the steady-state density matrix $\rho_ss$ in the local basis, as shown in Fig.~\ref{Fig2}(a). The results reveal that the steady state exhibits anomalously high overlaps with these two product states. $\ket{Z_2}=\ket{101010\cdots101010}$ and $\ket{Z_2'}=\ket{010101\cdots010101}$. 

We further examine this feature from a dynamic perspective by investigating the quantum fidelity, i.e., overlap between the density matrix at any time $\rho(t)$ and the initial time $\rho_0$ defined as follows \cite{fidelity} 

\begin{align}
F\left[\rho(t), \rho_0\right]=\operatorname{Tr}\left[\sqrt{\sqrt{\rho(t)} \rho_0 \sqrt{\rho(t)}}~\right]. \label{Fidelity}
\end{align}

Fig.\ref{Fig2} (b) plots the time evolution of quantum fidelity for two different initial states, which shows that the fidelity approaches zero rapidly and dynamical revival complete absence for the initial state without Rydberg excitations $\ket{g}=\ket{000\cdots000}$. In contrast, a remarkable periodic oscillation in time $\ket{Z_2}$ with period $T=2\pi/\Delta E$ (see the inset of Fig.\ref{Fig2}(b)) with the energy separation $\Delta E\approx1.33$ between adjacent QMBS near zero-energy. Moreover, the long-time steady value is non-zero, reflecting the structure in steady-state distribution on the local basis, as shown in Fig.\ref{Fig2}(a). It is worth pointing out that the conclusion above is far from trivial. This is because the jump operators we required can be principally constructed from non-Hermitian ones that make the steady state in the form $\rho_{ss}=\ket{\phi_n}\bra{\phi_n}$ for any eigenstates $\ket{\phi_n}$ of the system Hamiltonian. This kind of steady state is nothing but the dark state, i.e., $O_j\ket{\phi_n}=0$ for any $j$. However, this generally requires the condition that the jump operator is non-local, which is usually unrealistic. More critically, to guarantee the above dark state as an eigenstate, one has to require the prior information of the eigenstate $\ket{\phi_n}$, which could be unfeasible. Thus, seeking out a local or quasi-local operator acting only on a few neighboring sites is more realistic and desirable, which is our important guideline. 

\begin{figure*}[ht]
	\includegraphics[width=5.6cm]{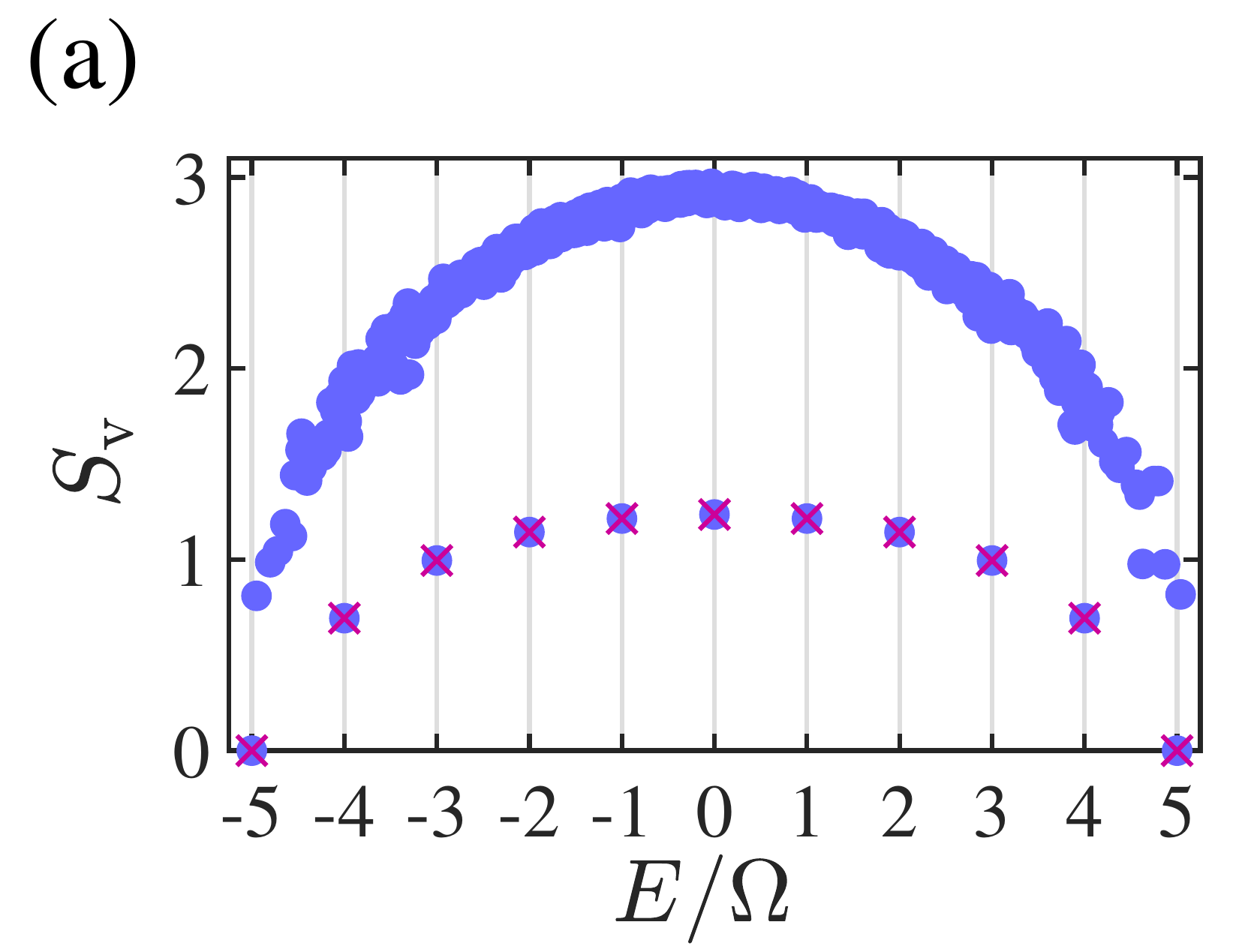}
	\includegraphics[width=5.6cm]{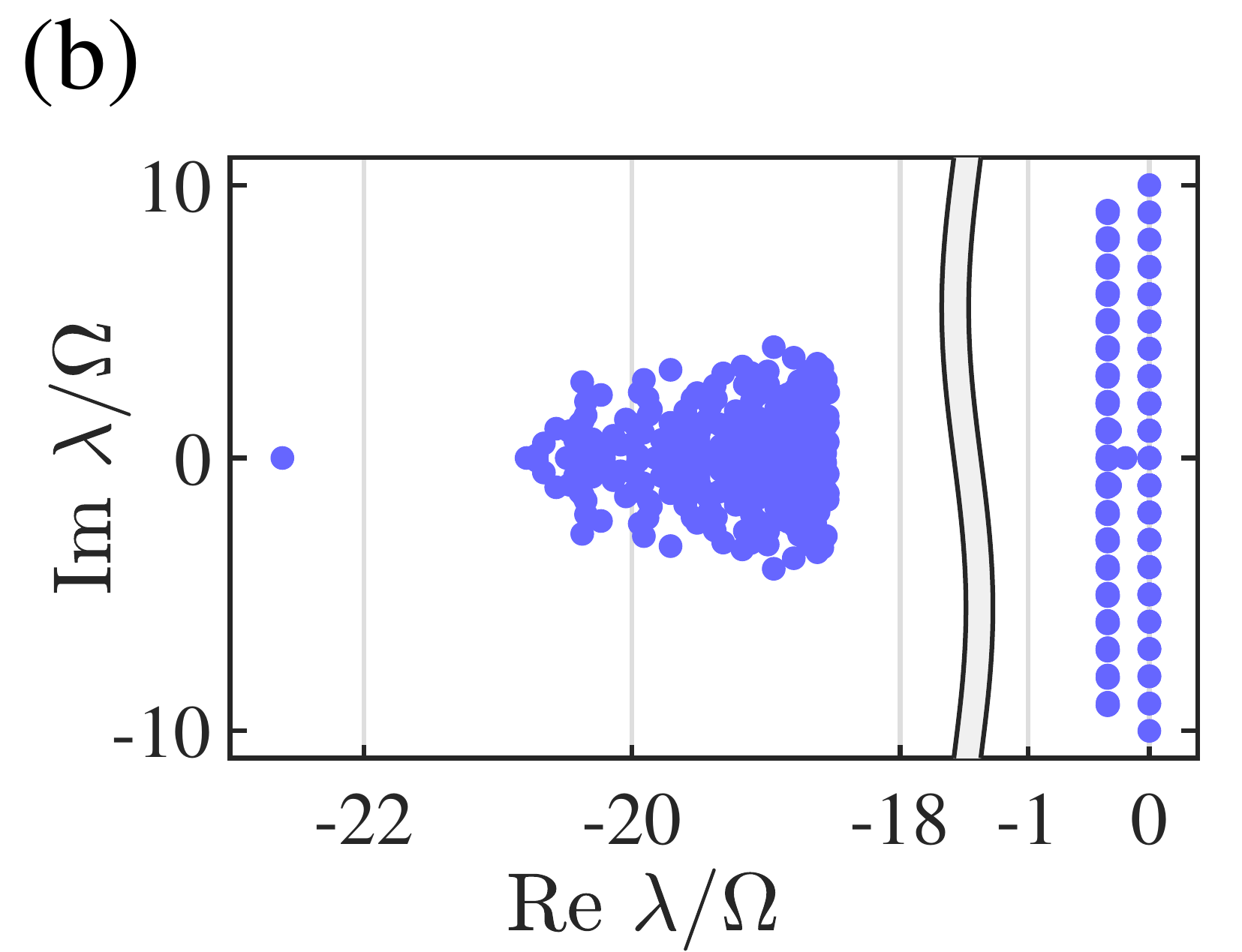}
	\includegraphics[width=5.6cm]{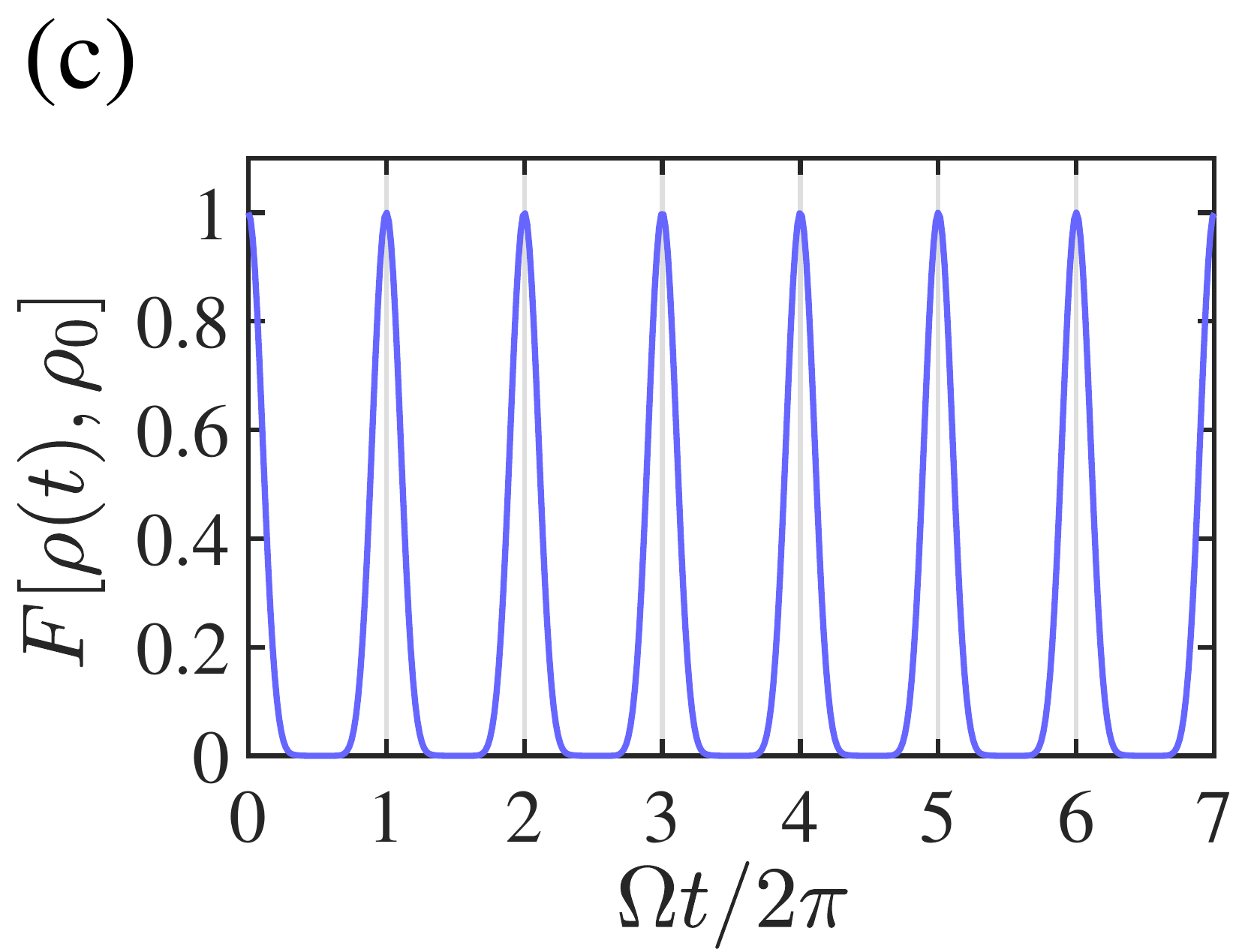}
	\caption{(a) The bipartite entanglement entropy of eigenstates for the toy model as a function of energy. Except for thermal states in the bulk, which have large entropies, a group of peculiar eigenstates with low entropies are nothing but QMBS. (b) The spectrum of the Liouvillian superoperator for the dissipative phase $\theta=0$ and dissipation strength $\gamma=0.1$. Here we just show eigenvalues with $200$ smallest and $200$ largest real parts. (c) Time evolution of quantum fidelity from the initial state $\rho_0=\ket{\phi_0}\bra{\phi_0}$ with $\ket{\phi_0}=\ket{\uparrow\uparrow\uparrow\cdots\uparrow\uparrow\uparrow}$ which shows a perfect revival with period $2\pi/\Omega$. Here we choose the system parameter $\Omega=1$, and coupling coefficients $J_{j-1, j+2}$ are chosen by independent normal distribution. The system size is $L=10$ with the periodic boundary condition.}
	\label{Fig4}
\end{figure*}

\textit{\color{blue}Dissipative toy model.} 
In addition to the PXP model studied above, our protocol can also apply to those systems with exactly solvable QMBS. To illustrate this, we use the toy model, which hosts exact QMBS as a concrete example whose Hamiltonian is written as \cite{Toymodel}

\begin{align}
H_{\mathrm{toy}}=\frac{\Omega}{2} \sum_j \sigma_j^x+\sum_j V_{j-1, j+2} P_{j, j+1}, \label{Ham_toy}
\end{align}
where $P_{j, j+1}=(1-\vec{\sigma}_j\cdot\vec{\sigma}_{j+1})/4$ represent the local projector on spin-singlet of a neighboring pair and $V_{i j}$ denotes an interaction operator acting on two spins $(i, j)$, which is generically expressed as $V_{i, j}=\sum_{\alpha \beta} J_{i j}^{\alpha \beta} \sigma_i^\alpha \sigma_j^\beta $ with arbitrary coupling coefficients $J_{i j}^{\alpha \beta}$ ($\alpha \beta=x,y,z$). This toy model is non-integrable due to the randomicity of interaction coefficients $J_{i j}$. We can also find that the Hamiltonian \eqref{Ham_toy} does not commute with the local projector $P_{j, j+1}$ and total magnetism $S^x=\sum_j\sigma_j^x/2$ and thereby there is no any apparent local symmetries. However, one can verify that states in the subspace $\ket{\mathcal{S}_n}=\ket{S=L/2, m_x=(n-L/2-1)}$ ($n=1,2,\dots,L+1$) with the total number of sites $L$, are eigenstates whose energies are exactly equidistant $E=\Omega m_x$. In contrast, other states outside of the subspace are affected by interaction terms $V_{j-1, j+2}$, which hybridize each other to form thermal eigenstates. The toy model belongs to embedding formalism \cite{Mori} where the QMBS are embedded into a generic non-integrable model. We will see that these exactly solvable states serve as QMBS and can also become steady states of a Liouvillian. We first numerically compute the bipartite entanglement entropy for all eigenstates of the toy model as shown in Fig.\ref{Fig4}(a) where some states with anomalously low entanglement and equally spaced energies are clearly visible while other states have large entanglement entropy. 
To investigate properties of QMBS under dissipation, we calculate the Liouvillian spectrum under the dissipation operators \eqref{Dissipator} at $\theta=0$,  where the zero-energy state (steady-state) is exactly composed of QMBS, i.e., $\rho_{ss}^n=\ket{\mathcal{S}_n}\bra{\mathcal{S}_n}$. A more striking consequence is that a string of states at equal spacing on the imaginary axis exists, which are non-decay modes because their real parts are zero. In fact, all non-decay modes originate from degenerate steady-state subspace where we can build up a series of independent non-decay modes of the Liouvillian superoperator by means of QMBS as $\ket{\mathcal{S}_m}\bra{\mathcal{S}_n}$ whose eigenvalue is $i(E_n-E_m)=i\Omega(n-m)$. The degenerate steady states and the associated equally spaced imaginary parts can contribute to nontrivial dissipative dynamics with perfect revival without decay, as shown in Fig.\ref{Fig4}(c). Note that the system does not require any symmetry to relaxes to steady-state with QMBS, which is different from previous studies with dynamical symmetries \cite{Buca2019,Booker2020,Buca2020,Chinzei2020,Guarnieri2022} 

To sum up, the use of non-local dissipators enables a system to be driven into a steady state predominantly composed of QMBS in both the PXP model and the toy model with exactly solvable QMBS. Furthermore, this approach is applicable to systems where QMBS emerge through spectrum-generating algebra (see Supplementary Material for further details).

\begin{figure}[H]
	\includegraphics[width=8.5cm]{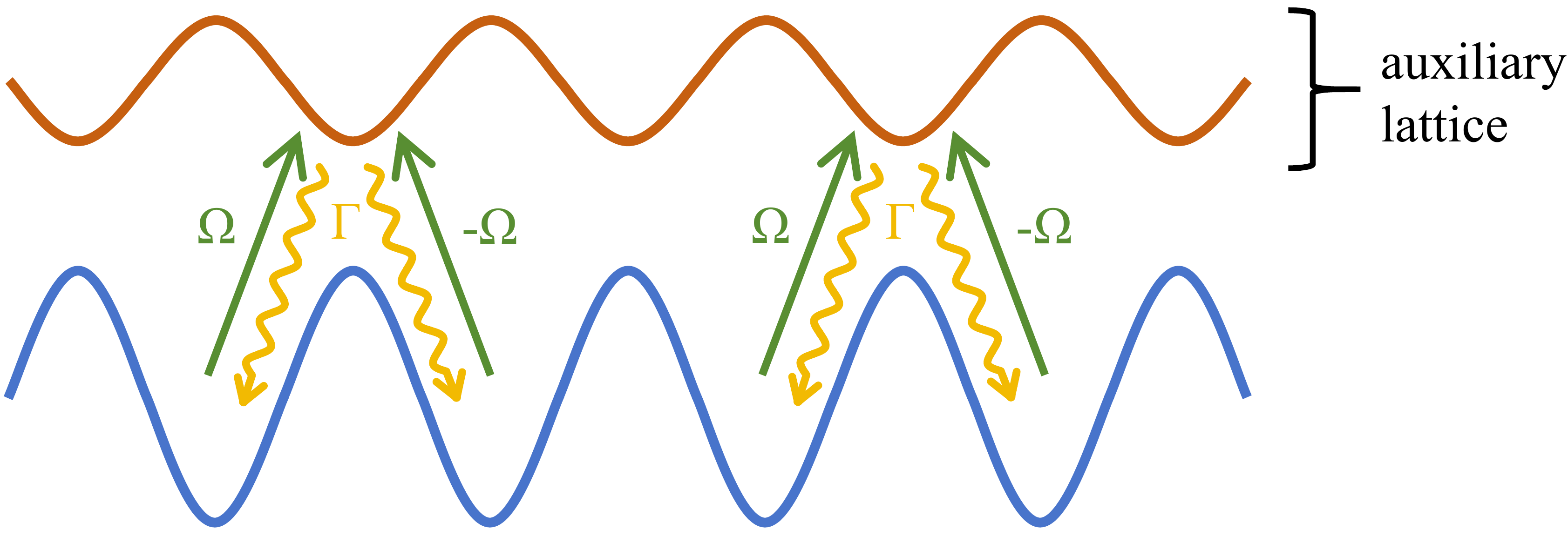}
	\caption{The schematic diagram for realizing dissipation operators \eqref{Dissipator} at $\theta=0$. The upper lattice represents the auxiliary lattice, while the lower lattice corresponds to the system lattice. Particles located at two adjacent sites in the system lattice are coupled to an intermediate auxiliary site through spatially modulated Raman laser fields, resulting in opposite couplings with Rabi frequencies $\pm \Omega$. This configuration effectively realizes the annihilation process $(\sigma_j-\sigma_{j+1})$. Subsequently, the creation process $\left(\sigma_j^{\dagger}+\sigma_{j+1}^{\dagger}\right)$ is achieved via isotropic spontaneous emission, wherein atoms decay at a rate $\Gamma$ and return to their original sites in the system lattice. This combined mechanism establishes a practical method for implementing the desired dissipation operators.
}
	\label{Fig_setup}
\end{figure}

\textit{\color{blue}Experimental realization.} As we pointed out the local or quasi-local operators are much easier to implement than those non-local ones. Here we provide an experimental scheme to realize our dissipation operators \eqref{Dissipator} concretely. First, we produce two lattice systems: system lattice and auxiliary lattice, as illustrated in Fig.\ref{Fig_setup}. Then, coherently couples two neighboring sites of the system to an auxiliary site between them with anti-symmetric couplings $\Omega$ and $-\Omega$ by spatial modulation of
the Raman laser \cite{Jump1}. In this way, the annihilation part of dissipation operators \eqref{Dissipator} is realized. Second, after the annihilation of
the anti-symmetric superposition of atoms (or spins) in the system lattice, the creation process sequentially is realized by spontaneous radiation where the atom decays back to the system sites from the auxiliary lattice. This creation process is symmetric since the spontaneous emission is isotropic.
In this way, one can realize the jump operator of the form implement
An arbitrary phase $\theta$ can be implemented via resonators array coupled by superconductive qubits \cite{Yusipov,BHchain}.
Further, we propose a quench dynamics scheme to measure the scarred nature of the steady state. Suppose the system has relaxed to the QMBS-dominated steady state under the action of quasi-local dissipation operators. Then we remove dissipation, and thereby, the system evolves under its Hamiltonian, where diagonal elements $\rho_{nn}$ of the density matrix in eigenbasis $\{\ket{\phi_n}\}$ remains unchanged and thereby the QMBS with associated scarred quench dynamics can be detected after removing dissipation.

\textit{\color{blue}Summary and outlook.} We have investigated the robustness of QMBS under the influence of dissipation. By employing quasi-local dissipation operators, we demonstrated that a system with QMBS can be driven into a steady state predominantly composed of QMBS. This conclusion is supported by our analysis of three distinct categories of QMBS. Our results reveal that dissipation can induce a weak ergodicity-breaking steady state, and the corresponding dissipative dynamics exhibit scarred relaxation with dynamical revivals.

Importantly, our proposed protocol is highly general, as the dissipation operators are model-independent, requiring no prior knowledge of the system's eigenstates. This universality mirrors the mechanism of dissipation-induced localization, where quasi-local dissipation stabilizes QMBS in a variety of systems. Beyond QMBS, other manifestations of weak ergodicity breaking, such as Hilbert space fragmentation, offer intriguing avenues for future exploration. Studying dissipation-induced non-thermal steady states in such contexts may significantly enhance our understanding of thermalization and ergodicity breaking in open quantum many-body systems.

Furthermore, we have discussed the experimental feasibility of realizing quasi-local dissipation operators in cold atom setups. Such implementations provide a pathway to making the signatures of ergodicity breaking observable in dissipative dynamics and open up exciting possibilities for experimentally stabilizing QMBS.


\section*{Acknowledgements}
We thank C. G. Liang for helpful discussions. The work is supported by National Natural Science Foundation of China (Grant No. 12304290 and Grant No. 12104205). WY acknowledges support from the National Key R\&D Program of China under Grant No.2022YFA1405800. LP also acknowledges support from the Fundamental Research Funds for the Central Universities. \\

%

\setcounter{equation}{0}
\setcounter{figure}{0}
\setcounter{table}{0}
\setcounter{section}{0}
\renewcommand{\theequation}{S\arabic{equation}}
\renewcommand{\thefigure}{S\arabic{figure}}
\renewcommand{\thesection}{S\arabic{section}}
\onecolumngrid
\flushbottom

\newpage

\subsection*{\normalsize Supplementary Material for ``Robustness of quantum many-body scars in the presence of Markovian baths"}\label{sec_supp}

\section{The QMBS and dissipative scarred dynamics in the model with conserved quantity} 
\begin{figure*}[!h]
	\includegraphics[width=8.0cm]{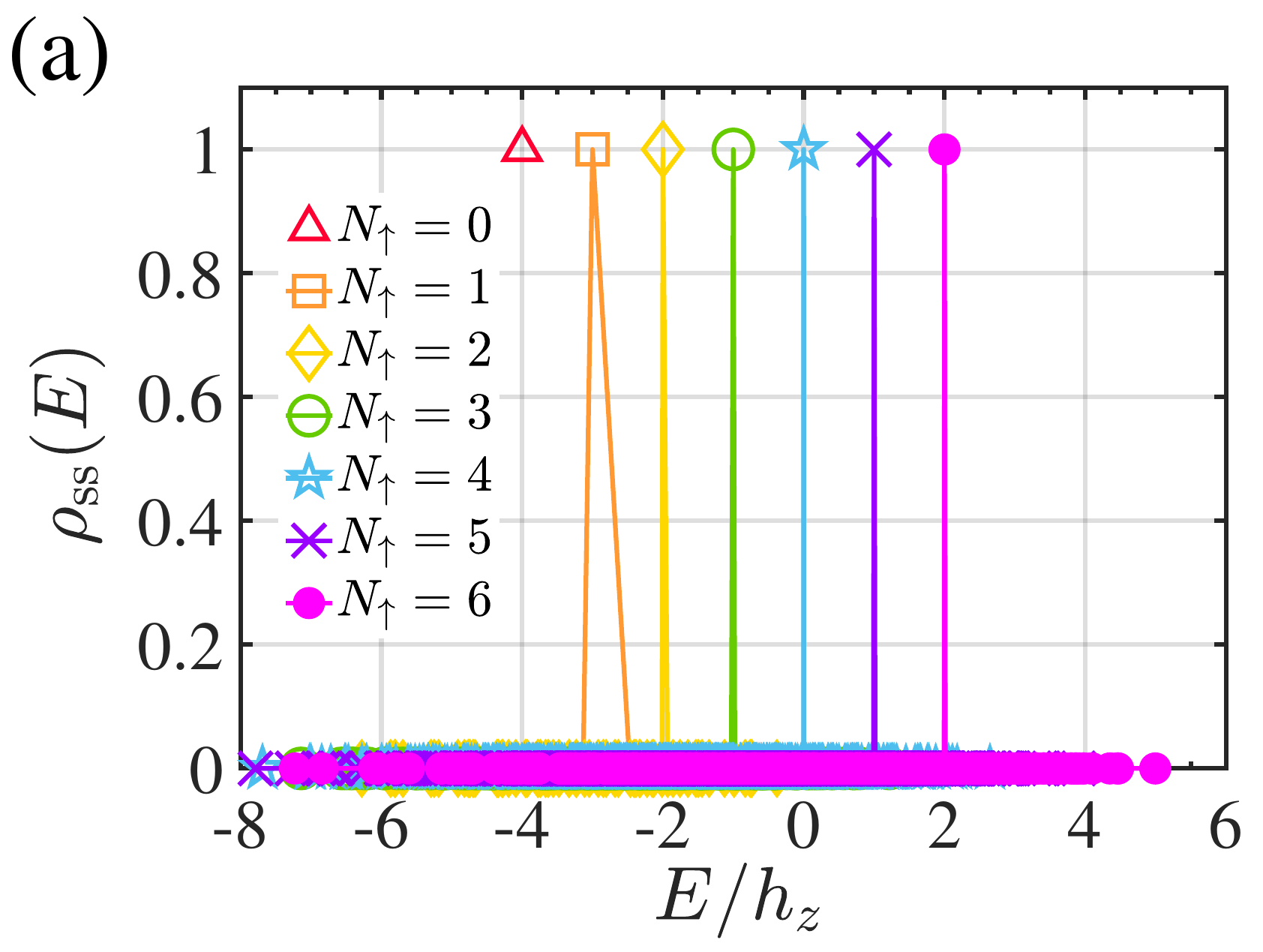}
	\includegraphics[width=8.0cm]{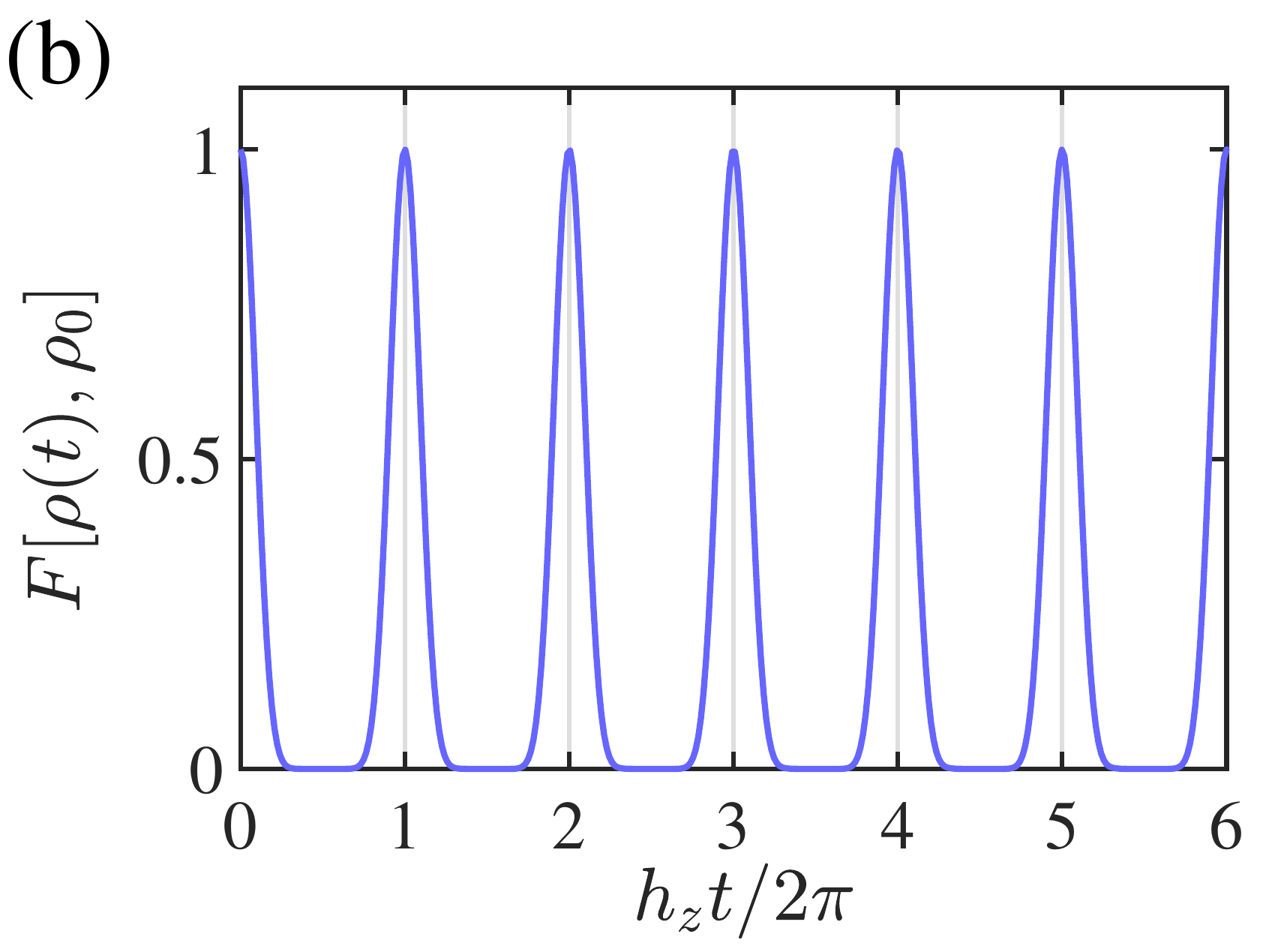}
	\caption{ (a) The distribution of the steady-state density matrix on eigenstate of the $J_1-J_2$ model in the eigenbasis of each subspace $\{N_{\uparrow},N_{\downarrow}\}$ with the dissipative phases $\theta=0$.
		(b) The dissipative dynamics of quantum fidelity from a ferromagnetic initial state$\ket{\phi_0}=\ket{\rightarrow\rightarrow\rightarrow\cdots\rightarrow\rightarrow\rightarrow}$. Here the parameters are chosen as $J_1=1,J_2=-0.2,D=1$, $h=0.5$ with system size $L=12$.  }
	\label{Fig_S1}
\end{figure*}

We have investigated the robustness of QMBS under the impact of dissipation. By choosing a kind of quasi-local dissipation operators, a system with QMBS can be driven to the steady state composed mainly by the QMBS. We confirm this by studying three categories of QMBS. Our results reveal that the dissipation can induce a weak ergodicity breaking steady-state, and the corresponding dissipative dynamics exhibit scarred relaxation with dynamical revival by suitably preparing an initial state. In this supplementary material, we apply our protocol to systems where the analytically solvable QMBS is equally spaced and can be constructed using spectrum-generating algebra. We will show that the QMBS can become exact steady states under the quasi-local dissipators introduced in the main text (see Eq.\eqref{O_j}). To show this, we take the $J_1-J_2$ model with Dzyaloshinskii-Moriya interaction (DMI) as an example whose Hamiltonian is
\begin{align}
	H_{J_1-J_2}=\sum_j\left(J_1 \mathbf{s}_j \cdot \mathbf{s}_{j+1}+J_2 \mathbf{s}_j \cdot \mathbf{s}_{j+2}\right)+H_{\mathrm{DMI}}+h_zS_{\text tot}^z,
\end{align}
where $H_{\mathrm{DMI}}=\sum_j \hat{\mathbf{z}} \cdot\left(\mathbf{s}_j \times \mathbf{s}_{j+1}\right)$ denotes the DMI and $S_{\text tot}=\sum_{j=1}^{L}s_{j}^{z}$. As discussed in Ref.\cite{J1_J2}, QMBS is composed of an equally spaced zero-momentum ($k=0$) magnon tower. Since the number of spin-up $N_{\uparrow}$ and spin-down $N_{\downarrow}$ is conserved, and also the dissipation operators do not change them. Hence, one can solve the Liouvillian superoperator and the steady-state individually in each subspace $\{N_{\uparrow},N_{\downarrow}\}$. As shown in Fig.\ref{Fig_S1}(a), the steady-state is the exact QMBS in each subspace, and the scar tower has equally spaced energy, which means it might be possible to generate dissipative dynamics with periodic revival. In fact, if we prepare a ferromagnetic state  $\ket{\rightarrow\rightarrow\rightarrow\cdots\rightarrow\rightarrow\rightarrow}$ as the initial state, which is an equally weighted superposition of QMBS ($k=0$ magnon), then the dissipative dynamics exhibits a perfect revival without decay, as shown in Fig.\ref{Fig_S1}(b).

\begin{thebibliography}{99}
\bibitem{Gogolin}  J. Eisert, M. Friesdorf, and C. Gogolin, Quantum many-body systems out of equilibrium, \href{https://doi.org/10.1038/nphys3215}{Nat. Phys. {\bf 7}, 124  (2015).}
\bibitem{Rigol2016} L. D' Alessio, Y. Kafri, A. Polkovnikov, and M. Rigol, From quantum chaos and eigenstate thermalization to statistical mechanics and thermodynamics, \href{https://doi.org/10.1080/00018732.2016.1198134}{Adv. Phys. {\bf 65}, 239 (2016).}

\bibitem{MBL1} R. Nandkishore and D. A. Huse, Many-body localization and thermalization in quantum statistical mechanics,
\href{https://doi.org/10.1146/annurev-conmatphys-031214-014726}{Annu. Rev. Condens. Matter Phys. {\bf 6}, 15 (2015).}
\bibitem{MBL2}  D. A. Abanin, E. Altman, I. Bloch, and M. Serbyn,
Colloquium: Many-body localization, thermalization,
and entanglement, \href{https://doi.org/10.1103/RevModPhys.91.021001}{Rev. Mod. Phys. {\bf 91}, 021001 (2019).}
	 
\bibitem{StrongETH1} B. L. Altshuler, Y. Gefen, A. Kamenev, and L. S. Levitov, Quasiparticle Lifetime in a Finite System: A Nonperturbative Approach, \href{http://dx.doi.org/10.1103/PhysRevLett.78.2803}{Phys. Rev. Lett. {\bf 78}, 2803 (1997).}
\bibitem{StrongETH2} D. Basko, I. Aleiner, and B. Altshuler, Metal–insulator transition in a weakly interacting many-electron system with localized single-particle states,  \href{http://dx.doi.org/10.1016/j.aop.2005.11.014}{Ann. Phys. (Amsterdam) {\bf 321}, 1126 (2006). }

\bibitem{StrongETH3} I. V. Gornyi, A. D. Mirlin,
and D. G. Polyakov, Interacting Electrons in Disordered Wires: Anderson Localization and Low-$T$ Transport,  \href{http://dx.doi.org/10.1103/PhysRevLett.95.206603}{Phys. Rev. Lett. {\bf 95}, 206603 (2005).}

\bibitem{StrongETH4} V. Oganesyan and D. A. Huse, Localization of interacting fermions at high temperature, \href{http://dx.doi.org/10.1103/PhysRevB.75.155111}{Phys. Rev. B {\bf 75}, 155111 (2007).}
	
\bibitem{StrongETH5} A. Pal and D. A. Huse, Many-body localization phase transition,  \href{http://dx.doi.org/10.1103/PhysRevB.82.174411}{Phys. Rev. B {\bf 82}, 174411 (2010).}

\bibitem{ETH1}  J. M. Deutsch, Quantum statistical mechanics in a closed system,  \href{https://journals.aps.org/pra/abstract/10.1103/PhysRevA.43.2046}{Phys. Rev. A {\bf 43}, 2046 (1991).}
\bibitem{ETH2} M. Srednicki, Chaos and quantum thermalization, \href{https://journals.aps.org/pre/abstract/10.1103/PhysRevE.50.888}{Phys. Rev. E {\bf 50}, 888 (1994).}
\bibitem{ETH3} M. Rigol, V. Dunjko, and M. Olshanii, \href{https://www.nature.com/articles/nature06838}{Nature (London) {\bf 452}, 854 (2008).}
 
\bibitem{OpenMBL_1} E. Levi, M. Heyl, I. Lesanovsky, and J. P. Garrahan, Robustness of Many-Body Localization in the Presence of Dissipation, \href{https://journals.aps.org/prl/abstract/10.1103/PhysRevLett.116.237203}{Phys. Rev. Lett. {\bf 116}, 237203 (2016).}	

\bibitem{OpenMBL_2} H. P. L\"uschen, P. Bordia, S. S. Hodgman, M. Schreiber, S. Sarkar, A. J. Daley, M. H. Fischer, E. Altman, I. Bloch, and U. Schneider, Signatures of Many-Body Localization in a Controlled Open Quantum System, \href{https://journals.aps.org/prx/abstract/10.1103/PhysRevX.7.011034}{Phys. Rev. X {\bf 7}, 011034 (2017).}	

\bibitem{OpenMBL_3} A. R.-Abadal, J. Choi, J. Zeiher, S. Hollerith, J. Rui, I. Bloch, and C. Gross, Many-Body Delocalization in the Presence of a Quantum Bath, \href{https://journals.aps.org/prx/abstract/10.1103/PhysRevX.9.041014}{Phys. Rev. X {\bf 9}, 041014 (2019).}	

\bibitem{Yusipov} I. Yusipov, T. Laptyeva, S. Denisov, and M. Ivanchenko, Localization in Open Quantum Systems, \href{https://journals.aps.org/prl/abstract/10.1103/PhysRevLett.118.070402}{Phys. Rev. Lett. {\bf 118}, 070402 (2017).} 

\bibitem{Yusipov2} I. Vakulchyk, I. Yusipov, M. Ivanchenko, S. Flach, and S. Denisov, Signatures of many-body localization in steady states of open quantum systems, \href{https://journals.aps.org/prb/abstract/10.1103/PhysRevB.98.020202}{Phys. Rev. B {\bf 98}, 020202(R) (2018).}

\bibitem{WYC_PRL} Y. Liu, Z. Wang, C. Yang, J. Jie, and Y. Wang, Dissipation-Induced Extended-Localized Transition, \href{https://journals.aps.org/prl/abstract/10.1103/PhysRevLett.132.216301}{Phys. Rev. Lett. {\bf 132}, 216301 (2024).}


\bibitem{WYC_MBL} Y. Hu, C. Yang, and Y. Wang, Can dissipation induce a transition between many-body localized and thermal states? 
\href{https://arxiv.org/abs/2407.13655}{arXiv:2407.13655 (2024).}

\bibitem{PL_mosaic} X.-P. Jiang, X. Yang, Y. Hu, and L. Pan, Dissipation induced ergodic-nonergodic transitions in finite-height mosaic Wannier-Stark lattices, \href{https://arxiv.org/abs/2407.17301}{arXiv:2407.17301 (2024).}

\bibitem{PL_3D} X. Yang, X.-P. Jiang, Z. Wei, Y. Wang, and L. Pan, Dissipation induced transition between extension and localization in the three-dimensional Anderson model, \href{https://arxiv.org/abs/2409.20319}{arXiv:2409.20319 (2024).}

\bibitem{scar1}  C. J. Turner, A. A. Michailidis, D. A. Abanin, M. Serbyn, and
Z. Papi{\'c}, Weak ergodicity breaking from quantum many-body
scars, \href{https://www.nature.com/articles/s41567-018-0137-5}{Nat. Phys. {\bf 14}, 745 (2018).}

\bibitem{scar2}  C. J. Turner, A. A. Michailidis, D. A. Abanin, M. Serbyn,and Z. Papi{\'c}, Quantum scarred eigenstates in a rydberg atom
chain: Entanglement, breakdown of thermalization, and stability
to perturbations, \href{https://www.nature.com/articles/s41567-018-0137-5}{Phys. Rev. B {\bf 98}, 155134 (2018).}

\bibitem{Fragment} S. Moudgalya and O. I. Motrunich, Hilbert space frag-
mentation and commutant algebras, \href{https://doi.org/10.1103/PhysRevX.12.011050}{Phys. Rev. X {\bf 12},
	011050 (2022).}
\bibitem{Buca2023}B. Bu\v{c}a, 
Unified Theory of Local Quantum Many-Body Dynamics: Eigenoperator Thermalization Theorems, \href{https://doi.org/10.1103/PhysRevX.13.031013}{Phys. Rev. X \textbf{13}, 031013 (2023).}

\bibitem{Rydberg_Exp1} H. Bernien, S. Schwartz, A. Keesling, H. Levine, A. Omran,
H. Pichler, S. Choi, A. S. Zibrov, M. Endres, M. Greiner, et al.,
Probing many-body dynamics on a 51-atom quantum simulator,
\href{https://www.nature.com/articles/nature24622}{Nature {\bf 551}, 579 (2017).}
\bibitem{Rydberg_Exp2} D. Bluvstein, A. Omran, H. Levine, A. Keesling, G. Semeghini, S. Ebadi, T. T. Wang, A. A. Michailidis, N.
Maskara, W. W. Ho, S. Choi, M. Serbyn, M. Greiner,
V. Vuleti{\'c}, and M. D. Lukin, Controlling quantum many-body dynamics in driven rydberg atom arrays, \href{https://www.science.org/doi/10.1126/science.abg2530}{Science {\bf 371}, 1355 (2021).}

\bibitem{scar_Review1} M. Serbyn, D. A. Abanin, and Z. Papi{\'c}, Quantum manybody
scars and weak breaking of ergodicity, \href{https://www.nature.com/articles/s41567-021-01230-2}{Nat. Phys. {\bf 17}, 675 (2021).}
\bibitem{scar_Review2}  S. Moudgalya, B. A. Bernevig, and N. Regnault, Quantum
many-body scars and hilbert space fragmentation: A review of
exact results,  \href{https://iopscience.iop.org/article/10.1088/1361-6633/ac73a0}{Rep. Prog. Phys. {\bf 85}, 086501 (2022).}
\bibitem{scar_Review3}  A. Chandran, T. Iadecola, V. Khemani, and R. Moessner, Quantum many-body scars: A quasiparticle perspective,
\href{https://www.annualreviews.org/content/journals/10.1146/annurev-conmatphys-031620-101617}{Annu. Rev. Condens. Matter Phys. {\bf 14}, 443 (2022).}





\bibitem{NH_scar1}  K. Pakrouski, P. N. Pallegar, F. K. Popov, and I. R. Klebanov, Group theoretic approach to many-body scar states
in fermionic lattice models, \href{https://journals.aps.org/prresearch/abstract/10.1103/PhysRevResearch.3.043156}{Phys. Rev. Research {\bf 3}, 043156
(2021).}
\bibitem{NH_scar2}  Q. Chen, S. A. Chen, and Z. Zhu, Weak ergodicity breaking in non-hermitian many-body systems, \href{https://www.scipost.org/10.21468/SciPostPhys.15.2.052?acad_field_slug=chemistry}{SciPost Phys. {\bf 15}, 052 (2023)}


\bibitem{Open_scar1}  H.-R. Wang, D. Yuan, S.-Y. Zhang, Z. Wang, D.-L. Deng, and L.-M. Duan, Embedding Quantum Many-Body Scars into Decoherence-Free Subspaces, \href{https://journals.aps.org/prl/abstract/10.1103/PhysRevLett.132.150401}{Phys. Rev. Lett. {\bf 132}, 150401 (2024).}

\bibitem{Open_scar2}  R. Shen, F. Qin, J.-Y. Desaules, Z. Papi{\'c}, and C. H. Lee, Enhanced Many-Body Quantum Scars from the Non-Hermitian Fock Skin Effect, \href{https://journals.aps.org/prl/abstract/10.1103/PhysRevLett.133.216601}{Phys. Rev. Lett. {\bf 133}, 216601 (2024).}





	\bibitem{Lindblad1} G. Lindblad, On the generators of quantum dynamical semigroups,
\href{https://link.springer.com/article/10.1007/BF01608499}{Commun. Math. Phys. {\bf 119}, 48 (1976).}

\bibitem{Lindblad2} V. Gorini, A. Kossakowski, and E. C. Sudarsahan,
Completely positive dynamical semigroups of N‐level systems,
\href{https://pubs.aip.org/aip/jmp/article/17/5/821/225427/Completely-positive-dynamical-semigroups-of-N}{J. Math. Phys. {\bf 17}, 821 (1976).}


\bibitem{Jump1} S. Diehl, A. Micheli, A. Kantian, B. Kraus, H. P. Büchler, and P. Zoller, Quantum states and phases in driven open quantum systems with cold atoms,  \href{https://www.nature.com/articles/nphys1073}{Nat. Phys. {\bf 4}, 878 (2008).}

\bibitem{Jump2} B. Kraus, H. P. Büchler, S. Diehl, A. Kantian, A. Micheli, and P. Zoller, Preparation of entangled states by quantum Markov processes,  \href{https://journals.aps.org/pra/abstract/10.1103/PhysRevA.78.042307}{Phys. Rev. A {\bf 78}, 042307 (2008).}

\bibitem{BHchain} D. Marcos, A. Tomadin, S. Diehl, and P. Rabl,
Photon condensation in circuit quantum electrodynamics by engineered dissipation, \href{https://iopscience.iop.org/article/10.1088/1367-2630/14/5/055005}{New J. Phys. {\bf 14}, 055005 (2012).}


\bibitem{Toymodel} W. Ho, S. Choi, H. Pichler, and M. D. Lukin, Periodic orbits, entanglement, and quantum many-body scars in
constrained models: Matrix product state approach, Phys.
\href{https://journals.aps.org/prl/abstract/10.1103/PhysRevLett.122.040603}{Rev. Lett. {\bf 122}, 040603 (2019).}

\bibitem{J1_J2} D. K. Mark and O. I. Motrunich, $\eta$-pairing states as true scars in an extended Hubbard model, \href{https://journals.aps.org/prb/abstract/10.1103/PhysRevB.102.075132}{Phys. Rev. B {\bf 102}, 075132 (2020).}

\bibitem{SGA} S. Moudgalya, N. Regnault1, and B. A. Bernevig, $\eta$-pairing in Hubbard models: From spectrum generating algebras to quantum many-body scars, \href{https://doi.org/10.1103/PhysRevB.102.085140}{Phys. Rev. B {\bf 102}, 085140 (2020).}

\bibitem{Liang} J. Ren, C. Liang, and C, Fang, Quasisymmetry Groups and Many-Body Scar Dynamics, \href{https://doi.org/10.1103/PhysRevLett.126.120604}{Phys. Rev. Lett. {\bf 126} 120604 (2021).}

\bibitem{Tang} L.-H. Tang, N. O'Dea, and A. Chandran, Multimagnon quantum many-body scars from tensor operators, \href{ttps://doi.org/10.1103/PhysRevResearch.4.043006
}{Phys. Rev. Research {\bf 4}, 043006 (2022).}

\bibitem{Popov} K. Pakrouski, P. N. Pallegar, F. K. Popov, I. R. Klebanov, Many Body Scars as a Group Invariant Sector of Hilbert Space, \href{ttps://doi.org/10.1103/PhysRevLett.125.230602}{Phys. Rev. Lett. {\bf 125}, 230602 (2020).}


\bibitem{blockade1} B. Olmos, M. Mller, and I. Lesanovsky, B. Olmos, M. Mller, and I. Lesanovsky, Thermalization of a
 strongly interacting 1d Rydberg lattice gas, \href{https://doi.org/10.1088/1367-2630/12/1/013024}{New J. Phys. 12, 013024 (2010).}

\bibitem{blockade2}I. Lesanovsky, B. Olmos, and J. P. Garrahan, Thermalization
in a Coherently Driven Ensemble of Two-Level Systems, \href{https://doi.org/10.1103/PhysRevLett.105.100603}{Phys. Rev. Lett. {\bf 105}, 100603 (2010).}

\bibitem{blockade3} C. Ates, J. P. Garrahan, and I. Lesanovsky, Thermalization
of a Strongly Interacting Closed Spin System: From Coherent Many-Body Dynamics to a Fokker-Planck Equation, \href{https://doi.org/10.1103/PhysRevLett.108.110603}{Phys. Rev. Lett. {\bf 108}, 110603 (2012).}

\bibitem{blockade4} S. Ji, C. Ates, J. P. Garrahan, and I. Lesanovsky, Equilibra-
tion of quantum hard rods in one dimension, \href{https://doi.org/10.1088/1742-5468/2013/02/P02005}{J. Stat. Mech. {\bf 2013}, P02005.}

\bibitem{Scar_perturbation1} C.-J. Lin, A. Chandran, and O. I. Motrunich, Slow thermalization of exact quantum many-body scar states under perturbations,\href{ https://doi.org/10.1103/PhysRevResearch.2.033044}{Phys. Rev. Research {\bf 2}, 033044 (2020).}

\bibitem{Scar_perturbation2} I. Mondragon-Shem, M. G. Vavilov, and I. Martin, Fate of Quantum Many-Body Scars in the Presence of Disorder,\href{ https://doi.org/10.1103/PRXQuantum.2.030349}{PRX Quantum {\bf 2}, 030349 (2021).}

\bibitem{fidelity} P. Zanardi, H. T. Quan, X. Wang, and C. P. Sun,
Mixed-state fidelity and quantum criticality at finite temperature,
\href{https://doi.org/10.1103/PhysRevA.75.032109}{Phys. Rev. A {\bf 75}, 032109 (2007).}

\bibitem{Mori} N. Shiraishi and T. Mori, Systematic construction of counterexamples to the eigenstate thermalization hypothesis, \href{https://doi.org/10.1103/PhysRevLett.119.030601}{Phys.
Rev. Lett. {\bf 119}, 030601 (2017).}


\bibitem{Buca2019} B. Bu\v{c}a, J. Tindall, and D. Jaksch, Non-stationary coherent quantum many-body dynamics through dissipation, \href{https://doi.org/10.1038/s41467-019-09757-y}{Nat. Commun. \textbf{10}, 1730 (2019).} 

\bibitem{Booker2020} C. Booker, B. Bu\v{c}a, and D. Jaksch, Non-stationarity and dissipative time crystals: spectral properties and finite-size effects, \href{https://doi.org/10.1088/1367-2630/ab9c2e}{New J. Phys. \textbf{22}, 085007 (2020).} 

\bibitem{Buca2020} B. Bu\v{c}a, C. Booker, M. Medenjak, and D. Jaksch, Bethe ansatz approach for dissipation: exact solutions of quantum many-body dynamics under loss, \href{https://doi.org/10.1088/1367-2630/abccbe}{New J. Phys. \textbf{22}, 123040 (2020).} 

\bibitem{Chinzei2020} K. Chinzei and T. N. Ikeda, Time crystals protected by Floquet dynamical symmetry in Hubbard models, \href{https://doi.org/10.1103/PhysRevLett.125.060601}{Phys. Rev. Lett. \textbf{125}, 060601 (2020).} 

\bibitem{Guarnieri2022} G. Guarnieri, M. T. Mitchison, A. Purkayastha, D. Jaksch, B. Bu\v{c}a, and J. Goold, Time periodicity from randomness in quantum systems, \href{https://doi.org/10.1103/PhysRevA.106.022209}{Phys. Rev. A \textbf{106}, 022209 (2022).}

\bibitem{Buca2022} Berislav Bu\v{c}a, Cameron Booker, Dieter Jaksch, Algebraic theory of quantum synchronization and limit cycles under dissipation, \href{https://doi.org/10.21468/SciPostPhys.12.3.097}{SciPost Phys. {\bf 12}, 097 (2022).}

\end{thebibliography}
\end{document}